\documentclass[aps,prl,reprint,groupedaddress]{revtex4-2}

\usepackage[colorlinks=true,citecolor=blue]{hyperref}

\usepackage{amsmath}
\usepackage{amssymb}
\usepackage{amsthm}
\usepackage{bbm}
\usepackage{verbatim}
\usepackage{graphicx}
\usepackage{color}

\newcommand{\sA}{\mathsf{A}}

\newcommand{\sD}{\mathsf{D}}
\newcommand{\sE}{\mathsf{E}}

\newcommand{\sH}{\mathsf{H}}

\newcommand{\sK}{\mathsf{K}}
\newcommand{\sL}{\mathsf{L}}

\newcommand{\sV}{\mathsf{V}}

\newcommand{\nrm}[1]{\lVert #1 \rVert}
\newcommand{\abs}[1]{| #1 |}

\newcommand{\tint}{\int_0^t \! dt' \;}
\newcommand{\fint}{\int_0^\infty \! dt \;}

\newcommand{\trans}{\mathsf{T}}

\begin{document}


\title{Dynamics of Micro and Nanoscale Systems in the Weak-Memory Regime}


\author{Kay Brandner}
\affiliation{
School of Physics and Astronomy, 
University of Nottingham, 
Nottingham, 
NG7 2RD, United Kingdom
}
\affiliation{
Centre for the Mathematics and Theoretical Physics of Quantum Non-Equilibrium Systems,
University of Nottingham, 
Nottingham, 
NG7 2RD, United Kingdom
}


\date{\today}

\begin{abstract}
Memory effects are ubiquitous in small-scale systems. 
They emerge from interactions between accessible and inaccessible degrees of freedom and give rise to evolution equations that are non-local in time. 
If the characteristic time scales of accessible and inaccessible degrees of freedom are sharply separated, locality can be restored through the standard Markov approximation. 
Here, we show that this approach can be rigorously extended to a well-defined weak-memory regime, where the relevant time scales can be of comparable order of magnitude. 
We derive explicit bounds on the error of the local approximation and a convergent perturbation scheme for its construction. 
Being applicable to any non-local time evolution equation that is autonomous and linear in the variables of interest, our theory provides a unifying framework for the systematic description of memory effects. 
\end{abstract}


\maketitle

Closed systems follow local time evolution laws, meaning their immediate future is fully determined by their present state. 
By contrast, the evolution of an open system at a given time may depend on its earlier states. 
Such memory effects arise from interactions with inaccessible degrees of freedom, which cannot be directly observed, and result in time non-local equations of motion. 
This phenomenon plays a particularly important role in micro and nanoscale systems, where typically only a fraction of the present dynamical degrees of freedom are accessible.
Brownian particles, for instance, drag along the invisible fluid molecules in their vicinity, thus altering their local environment in the future
\cite{berg-sorensen2005,jeney2008,franosch2011,narinder2018,ginot2022}. 
Similarly, the observable states of complex bio-molecules may have many internal configurations, which affect the visible dynamics but can often not be resolved in experiments \cite{hummer2015,martinez2019,hartich2021,ayaz2021,ertel2022,hartich2023,zhao2024}. 

In this article, we consider autonomous non-local time evolution equations with the general form \cite{fick1990,grabert1992,zwanzig2001}
\begin{equation}\label{eq:nleeq}
	\dot{x}_t = \sV x_t +  \tint \sK_{t'} x_{t-t'}. 
\end{equation}
Here, $x_t$ is a finite-dimensional vector, which characterizes the state of the open system, for instance in terms of a probability distribution over a discrete set of states.
The matrix $\sV$ describes the observable dynamics in the so-called adiabatic regime, where the inaccessible degrees of freedom relax to a unique stationary state virtually instantly \cite{esposito2012,bo2017,strasberg2019,seiferth2020}. 
We therefore refer to $\sV$ as the adiabatic generator. 
Any deviations from this limit are accounted for by the memory kernel $\sK_t$, which describes a retarded back-action of the inaccessible degrees of freedom on the accessible ones. 
Equations of the form \eqref{eq:nleeq} typically arise when inaccessible degrees of freedom are eliminated from linear time-local evolution equations on discrete state spaces; that is, for example, when time-homogeneous Markov jump networks are coarse grained by lumping subsets of micro-states into a finite number of meso-states \cite{landman1977,esposito2012,bo2017,strasberg2019,seiferth2020}, or when environments are eliminated from microscopic models of autonomous open quantum systems, as long as the Hilbert space of the accessible subsystem is finite-dimensional \cite{fick1990,grabert1992,zwanzig2001}.
This reduction can in general be achieved through projection operator methods, which are formally exact and can be systematically combined with expansions in small parameters \cite{fick1990,grabert1992,zwanzig2001}. 
Alternatively, time evolution equations of the type \eqref{eq:nleeq} can also emerge directly from stochastic modeling, for instance as ensemble-representations of semi-Markov jump processes \cite{feller1964,cinlar1975,breuer2009}, or as post-Markovian master equations for open quantum systems \cite{shabani2005,mazzola2010}.

Intuitively, it is clear that, close to the adiabatic regime, it should be possible to recover a local time evolution law, at least approximately and on some intermediate time scale. 
Indeed, a variety of techniques has been developed for this purpose, see, e.g., Refs.~\cite{nestmann2021a,nestmann2021,bruch2021,contreras-pulido2012,karlewski2014,
schaller2008,schaller2009,majenz2013,farina2019,mozgunov2020,
cattaneo2019,nathan2020,davidovic2020,kleinherbers2020,
trushechkin2021,davidovic2022}.
A central aim thereby is to find a time-homogeneous generator $\sL$ so that a solution of the local time evolution equation 
\begin{equation}\label{eq:leeq}
	\dot{y}_t = \sL y_t
\end{equation}
approximates the solution of the non-local equation \eqref{eq:nleeq}. 
The most common way to achieve this simplification is the Markov approximation, which averages over memory effects and leads to the generator \cite{fick1990,grabert1992}
\begin{equation}\label{eq:MG}
	\sL^1 = \sV + \fint \sK_t e^{-\sV t}. 
\end{equation}
Like most available methods, this approach relies on the assumption that the characteristic time scales of accessible and inaccessible degrees of freedom are sharply separated, i.e., differ by several orders of magnitude. 
However, the precise limits of this assumption are not entirely understood and universal methods to determine the generator $\sL$ to arbitrary accuracy are scarce and often limited to specific systems.
To help close these gaps, we seek to address two basic questions: 
Under what conditions and in what sense can a non-local time evolution equation be accurately approximated by a local one?
If such an approximation exists, is the corresponding generator unique and how can it be systematically constructed? 

Two remarks are in order before we proceed. 
First, dynamics with memory effects are commonly referred to as non-Markovian. 
However, the Eqs.~\eqref{eq:nleeq} and \eqref{eq:leeq} do not necessarily have to describe truly non-Markovian and Markovian stochastic processes, respectively \cite{vacchini2011b}. 
Since our results follow solely from the algebraic structure of these equations, we here use the more neutral terms non-local and local instead. 
Second, it is well known that locality in time can be restored with a time dependent generator \cite{tokuyama1975,tokuyama1976,grabert1977,grabert1978,chaturvedi1979,
shibata1980,shibata1977,chaturvedi1979,shibata1980,breuer2001}.
That is, Eq.~\eqref{eq:nleeq} is, under some technical conditions, formally equivalent to the time-convolutionless or TCL equation
\begin{equation}
	\dot{x}_t = \sL_t x_t.
\end{equation}
Our approach complements this method. 
Specifically, we will show that, within a well-defined weak-memory regime, Eq.~\eqref{eq:nleeq} is equivalent to the quasi-local time evolution equation 
\begin{equation}\label{eq:quleeq}
	\dot{x}_t = \sL x_t + \sE_t x_0.
\end{equation}
Here, the long-time dynamics of the system is described by the time independent generator $\sL$, to which the TCL generator converges in the long-time limit \cite{brandner2024}, and short-time corrections are accounted for by the  memory function $\sE_t$, which decays rapidly in time. 

We begin our analysis by collecting the relevant time scales of the problem at hand. 
The memory kernel $\sK_t$ is characterized by a magnitude $M$, which is determined by the strength of the interactions between the accessible and the inaccessible parts of the system, and a decay rate $k$, which is set by the relaxation time of the inaccessible degrees of freedom. 
For definiteness, we require that $\sK_t$ decays exponentially at long times such that 
\begin{equation}\label{eq:BndMK}
	\nrm{\sK_t}\leq M e^{-k t},
\end{equation}
where $\nrm{\cdot}$ is a matrix norm. 
This assumption is plausible, since we are interested in a regime where memory effects play a significant, yet subdominant, role. 
A third time scale is provided by the magnitude
\begin{equation}
	v = \nrm{\sV}
\end{equation}
of the adiabatic generator, which characterizes the free evolution of the accessible degrees of freedom. 

In the adiabatic limit $k\rightarrow\infty$, where $M$ is held fixed, any memory is wiped out and the solution of Eq.~\eqref{eq:nleeq} becomes $x_t = e^{\sL t} x_0$ with $\sL = \sV$. 
In general, however, the solution of Eq.~\eqref{eq:nleeq} cannot be expressed in terms of a simple matrix exponential, regardless of how the generator $\sL$ is chosen.
We therefore make the less restrictive ansatz
\begin{equation}\label{eq:MainT}
	x_t = e^{\sL t}\sA_t x_0 
		\quad\text{with}\quad \lim_{t\rightarrow\infty} \sA_t = \sD, 
		\quad\text{det}[\sD]\neq 0, 
\end{equation}
where the reduced propagator $\sA_t$ accounts for memory effects.
Requiring $\sA_t$ to approach a time-independent matrix $\sD$ makes it possible to asymptotically recover a local time evolution law. 
Specifically, the solution 
\begin{equation}\label{eq:LTA}
	y_t = e^{\sL t} \sD x_0
\end{equation}
of Eq.~\eqref{eq:leeq} provides a local approximation of $x_t$, which becomes arbitrarily accurate at long times if $\sA_t$ converges to $\sD$ sufficiently fast. 
The non-singularity of $\sD$, which is known as slippage matrix \cite{geigenmuller1983,haake1983,haake1985,suarez1992,gaspard1999,yu2000}, ensures the existence of a proper long-time approximation for any non-trivial initial state, i.e., $y_t\neq 0$ for any $x_0\neq 0$.  
It now remains to determine when a time-homogeneous generator $\sL$ with the above properties exists. 
This problem is addressed in Ref.~\cite{brandner2024}, where we show that, if the conditions 
\begin{equation}\label{eq:WMC}
	v<k \quad\text{and}\quad 4M < (k-v)^2
\end{equation}
are satisfied, $\sL$ can be chosen such that the relations \eqref{eq:MainT} and the bound
\begin{equation}\label{eq:BndLTA}
	\abs{x_t -y_t} \leq \frac{k-\eta}{\eta-\rho}\abs{x_0}e^{-\eta t}
\end{equation} 
hold. 
Here, $\abs{\cdot}$ can be any vector norm that is consistent with the matrix norm $\nrm{\cdot}$ and the parameters $\rho$ and $\eta$ are defined as 
\begin{align}
\label{eq:RhoEtaDef}
	\rho & = \frac{k+v-\sqrt{(k-v)^2-4M}}{2},\\
	\eta & = \frac{k+v+\sqrt{(k-v)^2-4M}}{2}.
	\nonumber
\end{align}
The bound \eqref{eq:BndMK} and the conditions \eqref{eq:WMC}  define the weak-memory regime. 
These conditions still require the memory time $1/k$ to be the shortest  relevant time scale. 
However, in contrast to the standard Markov approximation, which usually requires $v,\sqrt{M}\ll k$, they allow $v$, $\sqrt{M}$ and $k$ to be of the same order of magnitude.

The existence of the product representation \eqref{eq:MainT} of the solution of Eq.~\eqref{eq:nleeq} and the bound \eqref{eq:BndLTA} on the error of the long-time approximation \eqref{eq:LTA} are the first main results of this article. 
Instead of delving into the technicalities of their formal proof, which can be found in Ref.~\cite{brandner2024}, we illustrate these relations with a simple example. 
We choose the adiabatic generator and the memory kernel as
\begin{equation}\label{eq:Model1}
	\sV= v\sH, \;\;
	\sK= M\sH e^{-kt} \;\;\text{with}\;\;
	\sH= \frac{1}{2}\left[
			\begin{array}{rr}
				-1 & 1 \\ 1 &-1
			\end{array}\right].
\end{equation}
Here, $v, M, k>0$ are free parameters and $\nrm{\sH}_1 =1$, where $\nrm{\cdot}_1$ denotes the maximum absolute column sum norm
\footnote{
For $v<2k$ and $8M<(2k-v)^2$, the model \eqref{eq:Model1} describes a semi-Markov process with waiting time distribution 
\begin{equation*}
	\psi_t = \frac{(v\mu_+-vk-M)e^{-\mu_+ t}
		+ (vk-v\mu_- +M)e^{-\mu_- t}}{2(\mu_+-\mu_-)},
\end{equation*}
where $\mu_\pm = \bigl(2k+v\pm\sqrt{(2k-v)^2-8M}\bigr)/4$. 
}.
The time evolution equation \eqref{eq:nleeq} can then be interpreted, for example, as a simple model of a quantum dot in contact with a finite size reservoir \cite{schaller2014,brange2018,moreira2023}.
In this case, the state vector $x_t = [p^0_t, p^1_t]^\trans$ contains the probabilities $p^0_t$ and $p^1_t$ for the dot to be empty or occupied by a single charge.
The exact solution of Eq.~\eqref{eq:nleeq}, which can be easily found by Laplace transformation, has the form $x_t = e^{\sL t}\sA_t x_0$ with 
\begin{equation}
	\sL = \rho\sH \quad\text{and}\quad
	\sA_t = 1 + \frac{(e^{-(\eta-\rho)t}-1)(k-\eta)}{\eta-\rho}\sH,
\end{equation}
where $\rho$ and $\eta$ are defined as in Eq.~\eqref{eq:RhoEtaDef}. 
As long as $4M<(k-v)^2$, we have $\eta >\rho$ and $\sA_t$ converges to the non-singular matrix $\sD = 1 -(k-\eta)\sH/(\eta-\rho)$ in the limit $t\rightarrow\infty$. 
The corresponding long-time approximation \eqref{eq:LTA} then satisfies 
\begin{equation}
	\abs{x_t - y_t}_1 =
		\frac{k-\eta}{\eta-\rho}\abs{\sH x_0}_1 e^{-\eta t}.
\end{equation}
Hence, for $x_0=[1,0]^\trans$ and $x_0=[0,1]^\trans$, the bound \eqref{eq:BndLTA} is exactly saturated. 
Notably, for $4M=(k-v)^2$, the reduced propagator becomes $\sA_t = 1 - (k-\eta)\sH t$, and thus unbounded. 
For $4M>(k-v)^2$, we have $\eta -\rho = i \abs{\eta -\rho}$ and $\sA_t$ oscillates indefinitely. 
In neither of these cases can the generator $\sL$ be chosen such that $\sA_t$ converges to a non-singular matrix $\sD$, which shows that the condition $4M<(k-v)^2$ is generally necessary for the existence of the factorization \eqref{eq:MainT}; 
similar examples show that the same holds true for the condition $v<k$ \cite{brandner2024}.

In the above case study, we have extracted the generator and the reduced propagator from the exact solution of Eq.~\eqref{eq:nleeq}. 
In general, however, these objects cannot be found exactly. 
It is then still possible to determine them approximately through a systematic perturbation theory, which can be developed from the memory function $\sE_t$ that appears in Eq.~\eqref{eq:quleeq}. 
To this end, we equate the formal solutions of the Eqs.~\eqref{eq:nleeq} and \eqref{eq:quleeq} in Laplace space. 
Upon solving for the Laplace transform of the memory function and returning to the time domain, we find that $\sE_t$ solves the initial value problem 
\begin{align}
	\label{eq:ivp}
	\dot{\sE}_t & = \sK_t + \sE_t\sV + \tint \sE_{t'} \sK_{t-t'},\\
	\sE_0       & = \sV - \sL. 
	\nonumber
\end{align}
The solution of this problem is unique for any given $\sL$. 
However, the generator is yet unknown. 
This problem can be addressed as follows. 
We first switch to the dimensionless time $s=kt$ and introduce the rescaled variables $\bar{\sE}_s = \sE_{s/k}/k$, $\bar{\sV}=\sV/k$ and $\bar{\sK}_s = \sK_{s/k}/M$.  
The integro-differential equation \eqref{eq:ivp} then takes the form 
\begin{equation}\label{eq:MFResc}
	\frac{d}{ds}\bar{\sE}_s = \varphi\bar{\sK}_s 
		+ \bar{\sE}_s\bar{\sV} + \varphi \int_0^s \! ds' \; \bar{\sE}_{s'}
			\bar{\sK}_{s-s'},
\end{equation}
where $\varphi = M/k^2$ is a dimensionless parameter, which quantifies the memory strength.  
We now make the ansatz 
\begin{equation}
	\bar{\sE}_s = \sum_{n=1}^\infty \varphi^n \bar{\sE}^{(n)}_s.
\end{equation}
Inserting this expansion into Eq.~\eqref{eq:MFResc} and collecting terms of the same order in $\varphi$ yields a hierarchy of ordinary differential equations, whose formal solutions are 
\begin{align}
	\bar{\sE}_s^{(n)} & = \bar{\sE}^{(n)}_0 e^{\bar{\sV}s}
							+ \int_0^s \! ds'\int_0^{s'} \! ds'' \; \bar{\sE}^{(n-1)}_{s''}\bar{\sK}_{s'-s''} e^{\bar{\sV}(s-s')},
	\nonumber\\
	\bar{\sE}_s^{(1)} & = \bar{\sE}^{(1)}_0 e^{\bar{\sV}s} + \int_0^s \! ds' \; \bar{\sK}_{s'} e^{\bar{\sV}(s-s')}.
\end{align}
Since the memory function should vanish rapidly at long times, we choose the initial values $\bar{\sE}_0$ so that
\begin{equation}
 	\lim_{s\rightarrow\infty} \bar{\sE}_s^{(n)} e^{-\bar{\sV}s} =0.
\end{equation}
After returning to natural units, we thus obtain the recursion relations 
\begin{align}
\label{eq:RecE}
\sE^{(n)}_t & = -\int_t^\infty \! dt' \int_0^{t'}\! dt'' \; 
	\sE^{(n-1)}_{t''}\sK_{t'-t''}e^{\sV (t-t')},\\
\sE^{(1)}_t & = -\int_t^\infty \! dt' \; \sK_{t'} e^{\sV (t-t')},
\nonumber
\end{align}
which require only the adiabatic generator $\sV$ and the memory kernel $\sK_t$ as input. 
It thus becomes possible to iteratively calculate the  $n^\text{th}$-order approximation 
\begin{equation}\label{eq:EApprox}
	\sE^n_t = \sum_{m=1}^n \sE^{(m)}_t
\end{equation}
of the memory function. 
Once the memory function has been determined to sufficient accuracy, the corresponding approximations of the generator and the reduced propagator can  be obtained from the formulas
\begin{align}
	\label{eq:Approx}
	\sL^n & = \sV - \sE^n_0,\\
	\sA_t^n & = \mathsf{1} + \tint e^{-\sL^n t'} \sE_{t'}^n,
	\nonumber
\end{align}
which follow directly from Eq.~\eqref{eq:ivp} and the observation that $\sA_t$ and $\sE_t$ have to be connected through the relation $\sE_t = e^{\sL t}\dot{\sA}_t$ for the factorization \eqref{eq:MainT} to be consistent with the differential equation \eqref{eq:quleeq}. 
Notably, we recover the Markov generator \eqref{eq:MG} in first order with respect to $\varphi$. 

Up to this point, our discussion has been purely formal, since we have not paid any attention to convergence conditions. 
However, it is possible to put the above derivation on mathematically rigorous grounds \cite{brandner2024}.
As our second main result, we find that the successive approximations \eqref{eq:Approx} of the generator and the reduced propagator converge uniformly  on the open time interval $[0,\infty)$ in the limit $n\rightarrow\infty$, as long as the weak-memory conditions \eqref{eq:BndMK} and \eqref{eq:WMC} are satisfied. 
We have thus developed a systematic perturbation theory for the non-local time evolution equation \eqref{eq:nleeq}, where the memory strength $\varphi=M/k^2$ plays the role of a small parameter. 
Specifically, the $n^\mathrm{th}$-order approximation of $x_t$ takes the form 
\begin{equation}
	x^n_t = e^{\sL^n t} \sA^n_t.
\end{equation}
In the limit $n\rightarrow\infty$, this sequence converges uniformly to the exact solution of Eq.~\eqref{eq:nleeq} on any finite time interval $[0,t]$, for details, see Ref.~\cite{brandner2024}.

\begin{figure}
\includegraphics[width=8.5cm]{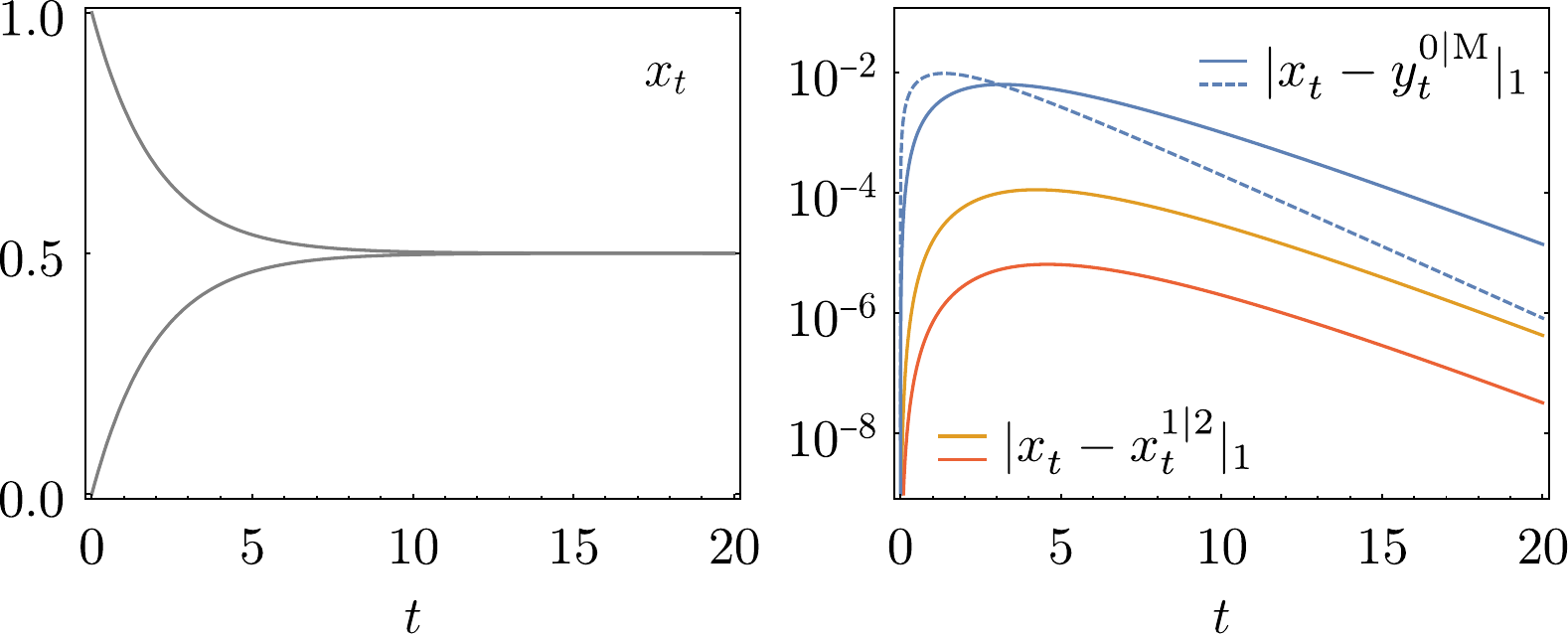}
\caption{\label{Fig}
Perturbation theory in the weak memory regime. 
The plots corresponds to the model of Eq.~\eqref{eq:Model1} for the initial state $x_0=[0,1]^\trans$. 
We have set $k=1$ so that all quantities become dimensionless. 
The remaining parameters are $v=0.5$ and $M=\varphi =0.01$ such that $4M/(k-v)^2 = 0.16$ and the weak-memory conditions \eqref{eq:WMC} are satisfied.
\textbf{Left:} Components of the exact solution $x_t$ of Eq.~\eqref{eq:nleeq}. 
\textbf{Right:} The orange and red lines show the error of the first and second order approximations, $x^1_t$ and $x^2_t$. 
For comparison, the solid and dashed blue lines shows the error of the adiabatic and Markov approximations, $y^0_t = e^{\sV t} x_0$ and $y^\mathrm{M}_t = e^{\sL^1 t} x_0$, respectively.}
\end{figure}

To illustrate this result, we revisit the model of Eq.~\eqref{eq:Model1}, for which the first and second order contributions to the memory function are given by 
\begin{align}
	\sE^{(1)}_t & = - \frac{M}{k-v}\sH e^{-k t},\\
	\sE^{(2)}_t & = - \frac{M^2}{(k-v)^3}\sH e^{-k t} -\frac{M^2t}{(k-v)^2}\sH e^{-k t}.
		\nonumber
\end{align} 
Using the formulas \eqref{eq:Approx}, it is straightforward to find the first and the second order approximations $x^1_t$ and $x^2_t$ of the exact solution $x_t$ of Eq.~\eqref{eq:nleeq}.
The errors of these approximations are plotted in Fig.~\ref{Fig} for selected parameter values. 
At short times, both $x^1_t$ and $x^2_t$ are significantly more accurate than the standard Markov approximation, $y^\mathrm{M}_t=e^{\sL^1 t} x_0$, which neglects the reduced propagator. 
At long times, the error of $x^1_t$ is comparable to the error of $y^\mathrm{M}_t$, which is to be expected, since both of these approximations use the same generator. 
The error of the second order approximation $x^2_t$, however, is consistently smaller than the error of $x^1_t$ and $y^\text{M}_t$ by more than one order of magnitude for all relevant times. 

We now return to our general program. 
As we have seen, the conditions \eqref{eq:BndMK} and \eqref{eq:WMC} precisely define a weak-memory regime, where a local approximation of the non-local time evolution equation \eqref{eq:nleeq} exists. 
The bound \eqref{eq:BndLTA} further shows that this approximation becomes arbitrarily accurate at sufficiently long times and the recursion relations \eqref{eq:RecE}, together with the formulas \eqref{eq:Approx}, provide a systematic scheme to construct this approximation order by order in the memory strength $\varphi=M/k^2$. 
It thus remains to determine in what sense the generator of the local approximation is unique. 
We first note that the factorization condition \eqref{eq:MainT} alone is insufficient to establish uniqueness. 
In fact, there generally exists a continuous family of generators, all of which are connected by similarity transformations, such that the solution of Eq.~\eqref{eq:nleeq} acquires the form \eqref{eq:MainT} \cite{brandner2024}. 
It is however possible to formulate a uniqueness criterion in terms of the memory function. 
To this end, we denote by $\sE_t$ the proper memory function that is uniquely determined by the recursion relations \eqref{eq:RecE} and by $\sL=\sV-\sE_0$ the corresponding proper generator. 
Next, we observe that, for any alternative generator $\sL'\neq\sL$, it is possible to construct a unique memory function $\sE'_t \neq \sE_t$ by solving the initial value problem \eqref{eq:ivp} \cite{burton2005}.  
Our third main result is that the proper memory function $\sE_t$ and any of its approximations $\sE^n_t$, as defined in Eq.~\eqref{eq:EApprox}, satisfy the bound 
\begin{equation}
	\nrm{\sE_t}, \nrm{\sE_t^n} \leq (k-\eta)e^{-\eta t}. 
\end{equation}
Any alternative memory function $\sE'_t\neq \sE_t$, however, violates this bound. 
More precisely, we have 
\begin{equation}
	\limsup_{t\rightarrow\infty} \; \nrm{\sE'_t}e^{\sigma t}=\infty
\end{equation} 
for any $\sigma>\rho$ \cite{brandner2024}. 
Hence, $\sE'_t$ either does not vanish at all in the long-time limit or at least decays significantly slower than $\sE_t$, since $\eta >\rho$. 
The proper generator $\sL$ is therefore unique in that it corresponds to the fastest decaying memory function. 

This result completes the agenda of the present article. 
In summary, we have developed a general mathematical framework that makes it possible to construct a generator $\sL$ and a non-singular slippage matrix $\sD$ such that the solution of the local time evolution equation \eqref{eq:leeq} for the initial condition $y_0 = \sD x_0$ comes arbitrarily close to the solution of the non-local equation \eqref{eq:nleeq} at sufficiently long times. 
The limits of this framework are precisely defined by the weak memory conditions \eqref{eq:BndMK} and \eqref{eq:WMC}, which do not require a strong separation of time scales. 
Instead, they allow the three parameters $v$, $\sqrt{M}$ and $k$, which represent the characteristic time scale of the accessible part of the system, the coupling strength between accessible and inaccessible degrees of freedom and the typical relaxation time of the latter, to be of the same order of magnitude. 
A crucial prerequisite for our theory is the existence of an exponentially decaying bound on the memory kernel. 
If the latter decays only algebraically, memory effects are expected to play a dominant role and can no longer be described with the methods developed in this article.
Nonetheless, our approach goes beyond the standard Markov approximation, which we recover as the first order of our perturbation theory for the generator $\sL$ in the memory strength $\varphi=M/k^2$. 

On the conceptual side, our results contribute towards closing a long-standing gap in the theory of open systems. 
While it is well understood how irrelevant degrees of freedom can be systematically eliminated from local time evolution equations by means of projection operator techniques \cite{fick1990,grabert1992,zwanzig2001}, transitioning from the resulting non-local time evolution equations, which in many instances take the form of Eq.~\eqref{eq:nleeq}, back to time local equations often requires non-systematic approximations. 
We stress that systematic expansions of local generators have been developed in the more recent literature \cite{nestmann2021a,nestmann2021,bruch2021,contreras-pulido2012,karlewski2014}, which we discuss further in Ref.~\cite{brandner2024}. 
However, the convergence properties of these schemes are not fully understood. 
At least within the weak-memory regime, these methods can now be augmented with a convergent perturbation theory in the memory strength.
From a practical perspective, our results provide a universal basis to further explore the role of memory effects in a variety of areas ranging from bio-molecular systems to open quantum systems
\cite{trushechkin2022,klippenstein2021,lapolla2019,breuer2016,
hofling2013}. 
In particular, it will be interesting to explore whether the quasi-local time evolution equation \eqref{eq:quleeq} can be endowed with a consistent thermodynamic structure, which would make it possible to systematically incorporate moderate memory effects into the existing frameworks of stochastic and quantum thermodynamics \cite{strasberg2022,shiraishi2023}. 
We leave this problem, along with potential extensions of our theory to systems that are non-autonomous, i.e., driven by time dependent fields, or follow non-linear equations of motion, as an appealing subject for future research.


\begin{acknowledgments}
\emph{Acknowledgments.--}
The author gratefully acknowledges insightful discussions with Paul Nemec, who verified the mathematical results reported in this article and corrected an error in the initial version of the part concerned with the uniqueness of the proper local generator.
This work was supported by the Medical Research Council (Grants No. MR/S034714/1 and MR/Y003845/1) and the Engineering and Physical Sciences Research Council (Grant No. EP/V031201/1).
The author further acknowledges support from the University of Nottingham through a Nottingham Research Fellowship. 
\end{acknowledgments}

\end{document}